# Optimizing I/O for Big Array Analytics[*]

Yi Zhang
Duke University
yizhang@cs.duke.edu

Jun Yang
Duke University
junyang@cs.duke.edu

## ABSTRACT

Big array analytics is becoming indispensable in answering important scientific and business questions. Most analysis tasks consist of multiple steps, each making one or multiple passes over the arrays to be analyzed and generating intermediate results. In the big data setting, I/O optimization is a key to efficient analytics. In this paper, we develop a framework and techniques for capturing a broad range of analysis tasks expressible in nested-loop forms, representing them in a declarative way, and optimizing their I/O by identifying sharing opportunities. Experiment results show that our optimizer is capable of finding execution plans that exploit nontrivial I/O sharing opportunities with significant savings.

## 1. INTRODUCTION

Scientific and business decisions increasingly rely on the analysis of big array data, e.g., vectors, matrices. It is often impossible or uneconomical to fit such data entirely in memory, even after partitioning and distribution in a parallel or cluster setting. Besides the big input data, analysis can write out big intermediate and/or final results. Thus, big array analytics today are I/O-intensive, and I/O optimization is critical in achieving high overall performance. Furthermore, data analysis has become more sophisticated—it may use linear algebra instead of relational operations as building blocks, and each step may exhibit a complex, multi-pass access pattern over its input and output. Optimizing I/O in this setting is challenging.

**Example 1.** *Consider a program with two steps, a matrix addition and a matrix multiplication: C = A + B, E = CD. Suppose the matrices are stored on disk in* blocks*. The blocks here are a logical storage and access unit, usually large in size and not to be confused with physical disk blocks. The following C-style program describes the operations involved. In the following (and the rest of this paper),* **each array access** *(such as C[i, j] below)* **represents a block access***, not an element access.*

```
for (i=0; i<n1; ++i)
  for (k=0; k<n2; ++k)
    C[i,k] = A[i,k] + B[i,k]; // s1
```

[*]This work is supported by the NSF award IIS-0916027 and an Innovation Research Program Award from HP Labs.

```
for (i=0; i<n1; ++i)
  for (j=0; j<n3; ++j)
    for (k=0; k<n2; ++k)
      E[i,j] += C[i,k] * D[k,j]; // s2
```

*There are two statements in the program: $s_1$ and $s_2$. If we regard each array access as an I/O, A and B are both read once, C is written once and then read $n_3$ times, D is read $n_1$ times, and E is written $n_2$ times and read $n_2 - 1$ times.[1] However, it is not hard to find some I/O-saving opportunities:*

1. *E[i, j] in $s_2$ can be kept in memory until the innermost loop is done, and written once.*

2. *C[i, k] in $s_2$ can be read once and kept in memory for the innermost loop, if we make j the innermost loop.*

3. *If 2 is done, the two loop nests can be* merged *and C in $s_1$ does not have to be written to disk at all.*

4. *D[k, j] in $s_2$ can be read once and kept in memory for the innermost loop, if we make i the innermost loop.*

The above example illustrates an important optimization idea—*I/O sharing*. When data is accessed repeatedly within the same processing step or by multiple steps in a data-intensive application, sharing I/O—i.e., retaining data in memory to avoid subsequent I/O—can reduce the overall running time. Although the example is simple in nature, it already reflects some intricacies of the I/O sharing problem, as listed below.

*Memory requirement*  Each I/O sharing opportunity in Example 1 results in different amount of I/O savings and requires different amount of memory in order to keep certain array blocks in memory. The opportunity that results in the most I/O savings may require more memory than available. It is necessary to analyze both factors.

*Legality*  Some opportunities (2, 3 and 4) change the original execution order of statement instances, which may or may not preserve the semantics of the program.

*Incompatibility of I/O sharing opportunities*  Some opportunities conflict and cannot be applied together. For example, Opportunity 1 requires $k$ to be the innermost loop for $s_2$, Opportunity 2 and 3 require $j$ to be the innermost loop, and Opportunity 4 requires $i$ to be the innermost loop.

*Dependence on parameters*  The optimal solution depends on not only the operators involved, but also the input parameters, namely sizes of arrays and their blocks. In Example 1, in the special case $n_3 = 1$, $s_2$ is surrounded by essentially only two loops. In this case, Opportunity 1 does not contradict Opportunities 2 and 3 any more; they can all be realized by the transformed program in Figure 1(a). While this special-case solution is easy for a human to produce, the general case of $n_3 \geq 1$ is not. If $E[i, j]$ is pinned in memory for continuous self accumulation (Opportunity 1), then it is impossible to avoid writing $C$ (Opportunities 2 and 3) unless $n_3 = 1$. However, it is still possible to "merge" the two loop nests and save a single pass of reading $C$, as shown in Figure 1(b). This solution subsumes the one in Figure 1(a). For a human to devise such a solution is nontrivial and error-prone; we would rather achieve this optimization automatically. As we will show in this paper, our optimizer can indeed find a parameterized solution for the most general case automatically.

---

[1]The listed code is simplified. Statement $s_2$ should actually be:
```
if (k==0) E[i,j]  = C[i,k] * D[k,j];
else      E[i,j] += C[i,k] * D[k,j];
```

```
for (i=0; i<n1; ++i) {
  // init E[i,0] with 0 in memory
  for (k=0; k<n2; ++k) {
    // read A[i,k] and B[i,k]
    C[i,k]  = A[i,k] + B[i,k]; // s1
    // pipeline C[i,k] from s1 to s2
    // read D[k,0]
    E[i,0] += C[i,k] * D[k,0]; // s2
  }
  // write E[i,0]
}
```
(a) Special case of $n_3 = 1$.

```
for (i=0; i<n1; ++i) {
  for (j=0; j<n3; ++j) {
    // init E[i,j] with 0 in memory
    for (k=0; k<n2; ++k) {
      if (j == 0) {
        // read A[i,k] and B[i,k]
        C[i,k] = A[i,k] + B[i,k]; // s1
        // write C[i,k]
      }
      // read D[k,j]
      // pipeline C[i,k] if j==0
      // read C[i,k] if j>0
      E[i,j] += C[i,k] * D[k,j]; // s2
    }
    // write E[i,j]
  }
}
```
(b) General case of $n_3 \geq 1$.

**Figure 1: Transformed code for Example 1.**

Existing database and compiler techniques fall short of solving the I/O sharing problem in our setting. First, *a database-like, operator-based approach does not allow full-fledged inter-operator optimization*. With this approach, users can write their programs in terms of logical operators such as matrix addition and multiplication. The system can provide for each logical operator a variety of physical implementations, each corresponding to a particular way of structuring the loops that implement the operator. This approach does allow for some I/O sharing opportunities such as pipelining between operators. Although this approach has enjoyed tremendous success in relational data processing, it is not suited for the array- and loop-centric applications that we consider, because the operators in our case have a far wider variety of implementation alternatives with complex data access patterns governed by many parameters. When putting our operators together for co-optimization, they cannot be treated as black boxes but need to be "opened up" so that the optimizer can tweak their inner workings further.[2] Otherwise, even a program as simple as in Example 1 cannot be handled. For instance, a database-like approach may be able to find a pipelining opportunity for $C$ if $n_3 = 1$, but will not be able to exploit "partial" pipelining as in Figure 1(b) if $n_3 > 1$.

---

[2] One might wonder if the need to "open up" operators can be avoided by making them more fine-grained. However, a complex operation often cannot be represented simply by a tree of fine-grained operators; instead, loop constructs would be required, which traditional database optimization does not handle. Indeed, our approach offers ways to reason with loops.

Second, *traditional compiler techniques cannot solve our I/O sharing problem because they lack explicit control over data reuse*. The compiler community has developed a plethora of techniques for automatic optimization of data locality [24, 9, 4, 7]. Their traditional focus is minimizing the traffic between CPU cache and memory, while ours is minimizing disk I/O. Although similar at a first glance, the two problems are fundamentally different. Traffic between cache and memory is hardware-managed and has peculiarities such as cache associativity; therefore, optimization tends to be best-effort, and does not produce a program that controls data sharing precisely. Traffic between memory and disk, on the other hand, is completely under our control, making *precise* control and analysis possible for our approach. Nonetheless, we have found the *polyhedral model*, which has been applied in a number of compiler optimizations [17, 11, 12, 13], to be a viable foundation to build on because it admits higher-level program analysis.

**Contributions** In this paper, we present *RIOTShare*, for optimizing I/O of loop-centric data-intensive programs. Building on the polyhedral model, we develop a new framework for capturing the I/O patterns of a program that is high-level enough to allow automatic extraction and reasoning of the I/O patterns, yet not too high-level to impede optimization flexibility (as black-box operators do). With this framework, we develop an optimizer that considers a rich space of plans (transformed programs), and is able to accurately determine their legality, I/O costs, and memory requirements. The optimizer employs *Apriori*-like search algorithm to enumerate different combinations of sharing opportunities and to look for legal, I/O-efficient plans under memory constraints. We demonstrate the effectiveness and accuracy of our optimizer through experiments.

## 2. RELATED WORK

Database systems rely on the buffer pool mechanism for sharing common I/O. This approach is rather low-level, opportunistic, and extremely sensitive to timing and the replacement policy used. There has also been much work on proactive work sharing. QPipe [16] proposes an on-demand simultaneous pipelining paradigm for maximizing data and work sharing across concurrent queries. It detects overlapping scans at run time and exploit the sharing opportunities using *circular scans*. Cooperative scans [27] is based on a similar idea, but coordinates I/O sharing using an *active buffer manager* and a policy called *relevance*, which is shown to be more effective than circular scans. Both approaches fall under the category of execution-time optimization, which is different from the principled, systematic optimization developed in this paper. Multi-query optimization [21, 19] tries to match common subqueries so that query processing can be partially shared. The recent DataPath system [2] relaxes the condition of sharing by employing a data-centric, push-based approach.

The aforementioned database-like, operator-based approaches have limited applicability in statistical and scientific data analysis workloads for two reasons. First, analytical operations typically have much more complex, parameter-governed data access patterns than most sequential-scan database operators. To support optimization of these complex I/O patterns across operators, operators need to be "opened up" so that the optmizer can reason about I/O sharing. Second, support for user-defined operators implementing customized analytical algorithms is a must. The system can no longer base optimization solely on some built-in knowledge of a static list of (physical) operators. This optimizable extensibility requirement again calls for a representation upon which both user-defined and built-in operators can be reasoned.

The compiler community has been working on automatic locality (data reuse) optimization of programs for decades. Most of the efforts have been devoted to locality at the cache level; examples include [24, 9, 4]. Since the cache is hardware-managed and its behavior depends on the machine's runtime state, the optimizer does not have explicit control over the data reuse at that level. Cache associativity further complicates the problem and only admits heuristic, instead of exact, solutions. Our problem is to optimize locality at the memory level and the key difference is that memory is explicitly managed by software, allowing us to develop a precise optimizer.

Tiling [4, 7], also called blocking or chunking, is a common technique to increase data locality. We solve a different problem in this paper: we assume tiling is already done and try to share the I/O of tiles by restructuring the data access patterns. The coarse-grained tiling optimizer of PLuTo [7] aims to increase parallelism and locality simultaneously, by minimizing the maximum of all *reuse distances* in the input program. In contrast, we directly optimize the total amount of data reuse, because the memory-level (as opposed to cache-level) data reuse can be precisely characterized.

Some compiler optimization ideas have been successfully applied in database systems. For example, MonetDB/X100 [6] adopts *vector processing* in place of the traditional tuple-at-a-time paradigm to achieve high CPU efficiency. HIQUE [18] uses a set of highly efficient code templates to customize code generation during query evaluation. It abandons the CPU-unfriendly iterator model and takes advantage of existing compiler optimizations to achieve high in-memory execution efficiency. Our work also has close ties to the compiler field, but we attack a different problem at a higher level.

Our optimizer represents and reasons about I/O sharing opportunities using the polyhedral model. The polyhedral model dates back to the seminal work of Karp, Miller and Winograd on uniform recurrence equations [17]. Because of its power of algebraic abstraction and transformation expressiveness, it has gain traction in the compiler field on some important optimization problems [14, 7]. However, to the best of our knowledge, this is the first time that it is applied to the I/O sharing problem.

## 3. OVERVIEW

Figure 2 shows the architecture of RIOTShare. The input to the system is a representation of an input program whose I/O we want to optimize. The representation is based on the polyhedral model, further discussed in Section 4.1, and can capture loop nests with conditional statements. We require the unit of I/O to be logical *blocks*, which is a standard practice to increase locality and reduce I/O overhead. Since we focus on optimizing I/O, we care only about read and write accesses to these blocks; the actual in-memory computation on them is unimportant. To obtain the representation for the input program in the polyhedral model, we can start with the program written using a library of high-level operators (such as matrix addition and multiplication), where the polyhedral representations of their implementations are already provided and can be assembled into a presentation for the entire program. Alternatively, we can obtain this representation by analyzing user-supplied pseudo-code (such as in Example 1) or source code for a user-defined operator or program, using standard code analysis tools like Clan.[3] The details of this preprocessing step are beyond the scope of this paper.

The next step is to identify data dependences as well as individual I/O *sharing opportunities*. Basically, a sharing opportunity signifies a data reuse relationship between two statements in the program. Note that the two statements can be the same one, in which case a *self* sharing opportunity occurs. We capture both dependences and sharing opportunities precisely, down to the *instance* level, i.e., individual accesses to the same block (as opposed to statements operating on the same array). In Section 4.3, we show how to express dependences and sharing opportunities concisely in polyhedral forms, which avoid costly enumeration of individual accesses and enable optimization.

[3] http://www.cse.ohio-state.edu/~pouchet/software/pocc

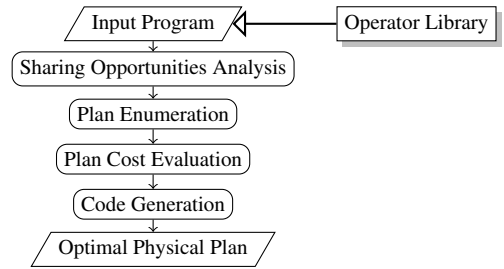

Figure 2: System architecture.

Given the dependences and sharing opportunities, our optimizer explores the space of plans (or schedules of data accesses) to find the I/O-optimal plan that is legal (i.e., respects all dependences) and meets the memory requirement. Intuitively, dependences and sharing opportunities translate to constraints on plans. The optimizer considers combinations of sharing opportunities using a strategy similar to the Apriori algorithm [1], to prune infeasible combinations of sharing opportunities. Legal plans are fed into a costing module, which evaluates their memory requirements and I/O costs. Finally, the best plan given the current avaialble memory resource is chosen and further converted into an executable plan. Section 5 discusses how the optimizer searches for and costs plans.

Finally, Section 6 demonstrates through experiments the effectiveness of our I/O sharing optimization framework. Section 7 concludes the paper and presents directions for future work.

## 4. A POLYHEDRAL I/O OPTIMIZATION FRAMEWORK

We first introduce some well-established concepts in the polyhedral model (Section 4.1), which serves as the foundation of our optimization framework. Next, we define the I/O sharing optimization problem (Section 4.2) and show how to characterize data dependences and I/O sharing opportunities (Section 4.3).

### 4.1 The Polyhedral Model

**Static-Control Programs** In this paper we focus on data-intensive programs that make "regular" accesses of out-of-core data. In particular, we assume the I/O patterns of the program can be described by a set of *static-control* loop nests and if conditionals, where the loop bounds, conditionals, and array access functions are *affine* combinations (linear combination plus a constant) of the enclosing loop variables and global parameters (e.g., array sizes). Note that this encompasses a large body of scientific and analytical programs, such as matrix addition, multiplication and factoriation, linear regression, table scans and nested loop joins in traditional databases, FILTER and FOREACH commands in Pig, etc. More general data flow programs, such as those with data-dependent control and non-affine conditionals, can also be cast into a static-control form by techniques like safe over-approximation [5].

**Iteration Domains** A program consists of a set $\mathcal{S}$ of *statements*. Each statement $s \in \mathcal{S}$ has an *iteration domain*, denoted $\mathbb{D}_s$, which describes the set of all executed *instances* of this statement. Each instance of $s$ is identified by the values of the loop variables surrounding $s$. In Example 1, $(i = 0, k = 0)$ is an instance of $s_1$, which is contained in its iteration domain $\mathbb{D}_{s_1} = \{(i, k) \in \mathbb{Z}^2 \mid 0 \leq i < n_1, 0 \leq k < n_2\}$. If $s$ is enclosed by a loop nest of depth $d_s$, then $\mathbb{D}_s$ is a parametric integer polyhedron [10] which is a subset of $\mathbb{Z}^{d_s}$ and

766

contains the statement instances as integer points. $\mathbb{D}_{s_1}$ is parameterized by $n_1$, $n_2$ and $n_3$. Geometrically, a polyhedron is a union of convex polyhera, each of which is the intersection of finitely many half-spaces and can be described by a system of linear inequalities. For example, $\mathbb{D}_{s_1}$ above can be written as a system of linear inequalities: $(i \geq 0) \wedge (-i + n_1 - 1 \geq 0) \wedge (k \geq 0) \wedge (-k + n_2 - 1 \geq 0)$, or equivalently in matrix form as:

$$\Delta_s \vec{x}_s = \Delta_s \begin{pmatrix} \vec{l}_s \\ \vec{p} \\ 1 \end{pmatrix} = \begin{pmatrix} 1 & 0 & 0 & 0 & 0 & 0 \\ -1 & 0 & 1 & 0 & 0 & -1 \\ 0 & 1 & 0 & 0 & 0 & 0 \\ 0 & -1 & 0 & 1 & 0 & -1 \end{pmatrix} \begin{pmatrix} i \\ k \\ n_1 \\ n_2 \\ n_3 \\ 1 \end{pmatrix} \geq \vec{0}.$$

We call $\vec{l}_s$ $s$'s loop *iteration vector*, $\vec{x}_s$ the *extended iteration vector*, and $\vec{p}$ the *parameter vector*. Different statements in a program can have (partly or completely) different iteration domains. For the brevity of presentation, we may drop the parameter vector and refer to the extended iteration vector simply as the iteration vector.

**Array Accesses** A statement $s$ can access multiple arrays. Each *access* is defined as a tuple $\mathfrak{a} = \langle s, t, A, \Phi \rangle$, where $s$ is the statement performing the access, $t \in \{R, W\}$ the type of access (read or write), $A$ the array accessed, and $\Phi$ a matrix describing the affine access function which maps $\vec{x}_s$ (the extended iteration vector of $s$) to a subscript in $A$. Each point in $A$'s subscript space corresponds to a block of array elements. $\Phi$ has as many rows as $A$'s dimensionality and as many columns as $\vec{x}_s$'s dimensionality. Note that $\Phi$ is required to uniquely identify an access because $s$ may access multiple parts of $A$. For example, there are three accesses in a statement $s$ `A[i,j]=A[i,j]+A[i,j]+A[i,j+1]`:

$$\left\langle s, W, A, \begin{pmatrix} 1 & 0 & 0 \\ 0 & 1 & 0 \end{pmatrix} \right\rangle, \left\langle s, R, A, \begin{pmatrix} 1 & 0 & 0 \\ 0 & 1 & 0 \end{pmatrix} \right\rangle, \left\langle s, R, A, \begin{pmatrix} 1 & 0 & 0 \\ 0 & 1 & 1 \end{pmatrix} \right\rangle.$$

`A[i,j]` and `A[i,j+1]` are regarded as different read accesses because they access different parts of $A$; the write (assignment) of `A[i,j]` is regarded as a different access from the reads of `A[i,j]` because of different access types. However, note the two reads of `A[i,j]` are treated as *one* access and have the same tuple representation, because they can always be serviced with only one I/O. In this paper, we assume each statement can have only one write access, which holds in most programming languages.

**Schedules** Each program has a schedule, which maps all statement instances in the program to an execution time. Formally, we define a *statement schedule* for statement $s$ to be an affine function (or matrix) $\Theta_s$ mapping $\mathbb{D}_s$, the iteration domain of $s$, to a *multidimensional time domain*, and a *program schedule* to be the set of all statement schedules in the program $\Theta = \{\Theta_s \mid s \in \mathcal{S}\}$. The time domain is a totally ordered set of vectors, where the order is lexicographic (the vector components can be thought of as, for example, year, month, day, etc.): $(x_1, \ldots, x_m) \prec (y_1, \ldots, y_m) \Leftrightarrow \exists r \in [1, m], (\forall i \in [1, r-1], x_i = y_i) \wedge (x_r < y_r)$. For the code in Example 1, one possible program schedule is (loop variables of $s_2$ are renamed to avoid confusion):

$$\Theta_{s_1}\vec{x}_{s_1} = \begin{pmatrix} 0 & 0 & 0 & 0 & 0 & 0 \\ 1 & 0 & 0 & 0 & 0 & 0 \\ 0 & 1 & 0 & 0 & 0 & 0 \end{pmatrix} \begin{pmatrix} i \\ k \\ n_1 \\ n_2 \\ n_3 \\ 1 \end{pmatrix} = \begin{pmatrix} 0 \\ i \\ k \end{pmatrix},$$

$$\Theta_{s_2}\vec{x}_{s_2} = \begin{pmatrix} 0 & 0 & 0 & 0 & 0 & 0 & 1 \\ 1 & 0 & 0 & 0 & 0 & 0 & 0 \\ 0 & 1 & 0 & 0 & 0 & 0 & 0 \\ 0 & 0 & 1 & 0 & 0 & 0 & 0 \end{pmatrix} \begin{pmatrix} i' \\ j' \\ k' \\ n_1 \\ n_2 \\ n_3 \\ 1 \end{pmatrix} = \begin{pmatrix} 1 \\ i' \\ j' \\ k' \end{pmatrix}.$$

Because of the 0 and 1 in the first component of the result time vectors, all instances of $s_1$ are scheduled before those of $s_2$. Also, $i$ appearing before $k$ in $\Theta_{s_1}$ corresponds to the fact that $i$ is an outer loop than $k$. Note that there are many equivalent schedules for the same program as far as the execution order is concerned. For example, changing $\Theta_{s_2}\vec{x}_{s_2}$ to $(2, i' + 1, j', k')$ does not change the relative execution order of statement instances in the program. Our optimizer works no matter which one of the equivalent schedules is specified for the input program.

### 4.2 Problem Definition

With the above preliminaries, the problem we tackle is the following. Given a memory cap and a static-control input program, whose iteration domains, array references, and original schedule are specified under the polyhedral model, find a *legal* transformation (represented by a schedule) of the given program such that I/O sharing is maximized (i.e., total I/O cost minimized) and memory consumption does not exceed the cap. A legal program schedule is one under which all data *dependences* in the original program schedule are observed; we formalize the notion of dependences in Section 4.3. We discuss how to compute I/O cost and memory consumption for a program schedule in Section 5.4.

**Why Explicitly Capping Memory?** We impose an explicit memory cap instead of relaxing the restraint and relying on the virtual memory mechanism. As verified in previous work such as [25], the virtual memory mechanism fails to utilize application-level memory usage information to optimally orchestrate contending consumers, and as a result, may lead to excessive paging for the types of programs we consider. Thus, we choose to impose a memory cap and control memory data reuse explicitly.

**Schedule Search Space** Recall that a (program) schedule is a set of affine functions, one for each statement, which maps the iteration instances to their scheduled execution time. It has been shown [13] that we can always find a schedule with dimension $\tilde{d} + 1$ for any static-control program, where $\tilde{d} = \max_{s \in \mathcal{S}} d_s$ and $d_s$ is the depth of the loop nest enclosing $s$ in the original program. Thus, we can safely restrict our search to $(\tilde{d} + 1)$-dimensional schedules only. Statement $s$'s schedule is then a list of $(\tilde{d} + 1)$ 1-d affine functions: $\Theta_s = (\theta_s^1, \ldots, \theta_s^{\tilde{d}+1})$, where each function corresponds to a row in the matrix and maps the iteration vector to a scalar time component.

As shown in [13], it is possible to make the last schedule dimension a constant, i.e., $\theta_s^{\tilde{d}+1} = c_s$, for all $s \in \mathcal{S}$, where $c_s$ denotes the textual position of $s$ in the transformed program under the schedule. For example, $\Theta_{s_1}\vec{x}_{s_1} = (k, 0, 1), \Theta_{s_2}\vec{x}_{s_2} = (i+n, j, 1), \Theta_{s_3}\vec{x}_{s_3} = (i+n, j, 2)$ is a schedule describing the program below:

```
for (k=0; k<n; ++k)
  // s1
for (i=0; i<n; ++i)
  for (j=0; j<n; ++j) {
    // s2 and s3
}
```

All instances of $s_1$ are scheduled before those of $s_2$ and $s_3$ due to the first schedule dimension. The order of the two instances of $s_2$ and $s_3$ within the same loop iteration are determined by the last constant dimension, which specifies the textual order of the two statements.

**Why the Polyhedral Model?** We choose the polyhedral model as the "language" for solving the I/O sharing problem for three reasons. First, it is well known that the polyhedral model captures a large space of transformations, such as loop interchange, reverse, skew, fusion, etc., and their compositions [12, 14, 7]. It can also handle programs with more general code than static-control loops [5].

Second, analysis of data flow in the polyhedral model is at the level of statement *instances* as opposed to just statements. This level

767

of detail makes it possible to build a precise cost-based optimizer that captures individual block accesses. For example, our optimizer is able to identify the "partial" sharing opportunity in Figure 1(b).

Third, the polyhedral model abstracts program transformations to make them amenable to automatic and systematic search. This is in direct contrast with traditional *syntactic* analysis, where programs are represented in abstract syntax trees and go through a series of pattern-matching and transformation steps, which does not lend itself to structured search [14].

### 4.3 Dependences and I/O Sharing Opportunities

Building on the polyhedral model, we show how to represent data dependences and I/O sharing opportunities, which are essential in determining the legality, I/O cost, and memory requirement of a program transformation (schedule). We begin with the notions of *co-accesses* and their *extent polyhedra*.

**Definition 1** (Co-Access and Extent Polyhedron). *Let $\mathcal{A}$ denote the set of all array block accesses in the program. A co-access, denoted $\mathfrak{a} \rightarrow \mathfrak{a}'$, is a pair of accesses in $\mathcal{A} \times \mathcal{A}$ to the same array; i.e., $\mathfrak{a}.A = \mathfrak{a}'.A$. The type of co-access $\mathfrak{a} \rightarrow \mathfrak{a}'$ is $\mathfrak{a}.t \rightarrow \mathfrak{a}'.t$, which is one of $R \rightarrow R$, $R \rightarrow W$, $W \rightarrow R$, and $W \rightarrow W$.*

*Suppose the original program schedule is $\Theta = \{\Theta_s \mid s \in \mathcal{S}\}$. The (extent) polyhedron of co-access $\mathfrak{a} \rightarrow \mathfrak{a}'$, where $\mathfrak{a} = \langle s, t, A, \Phi \rangle$ and $\mathfrak{a}' = \langle s', t', A, \Phi' \rangle$, is $\mathbb{P}(\mathfrak{a} \rightarrow \mathfrak{a}') = \{(\vec{x}, \vec{x}') \mid \vec{x} \in \mathbb{D}_s, \vec{x}' \in \mathbb{D}_{s'}, \Phi\vec{x} = \Phi'\vec{x}', \Theta_s\vec{x} < \Theta_{s'}\vec{x}'\}$.*

Thus, the extent polyhedron of a co-access is a polyhedron in the product space of the iteration domains of the two statements involved. Intuitively, it contains all pairs of statement instances that access the same array block ($\Phi\vec{x} = \Phi'\vec{x}'$), where the source instance executes before the target instance in the original schedule ($\Theta_s\vec{x} < \Theta_{s'}\vec{x}'$).

In the following, when no ambiguity exists (specifically, if statements $s$ and $s'$ each make only one access of a given type to the array $A$), we may omit $\Phi$ and $\Phi'$ and denote a co-access by $stA \rightarrow s't'A$.

Using the notion of co-accesses, we can now define data dependences and I/O sharing opportunities.

**Definition 2** (Dependences). *A (data) dependence is a co-access $\mathfrak{a} \rightarrow \mathfrak{a}'$ with type $R \rightarrow W$, $W \rightarrow R$, or $W \rightarrow W$ (i.e., at least one access is a write) and $\mathbb{P}(\mathfrak{a} \rightarrow \mathfrak{a}') \neq \emptyset$. Let $\mathcal{D}$ denote the set of all dependences in the original program.*

Intuitively, the polyhedron for a dependence specifies all data dependences among individual array block accesses. Given a dependence, for any pair of statement instances $(\vec{x}, \vec{x}')$ in its polyhedron, $\vec{x}$ *must* execute before $\vec{x}'$ under any legal schedule. Note that $R \rightarrow R$ co-accesses are not dependences because exchanging the order of two reads by itself does not affect program semantics.

I/O sharing opportunities are also defined using co-accesses, but they are different from dependences in subtle yet important ways.

**Definition 3** (Sharing Opportunities). *An I/O sharing opportunity is a co-access $\mathfrak{a} \rightarrow \mathfrak{a}'$ with type $W \rightarrow R$, $W \rightarrow W$, or $R \rightarrow R$ and $\mathbb{P}(\mathfrak{a} \rightarrow \mathfrak{a}') \neq \emptyset$. Let $O$ denote the set of all sharing opportunities in the original program.*

Intuitively, the polyhedron for a sharing opportunity identifies all possibilities for sharing I/O among individual array block accesses. Given a sharing opportunity, for any pair of statement instances $(\vec{x}, \vec{x}')$ in its polyhedron, $\vec{x}$ and $\vec{x}'$ may share I/O in accessing the same array block (at $\Phi\vec{x} = \Phi'\vec{x}'$). Note that a sharing opportunity merely indicates the potential for sharing but does not guarantee it; whether the potential is realized depends on the final schedule chosen. Specifically, depending on the type of the opportunity, sharing may happen in the following ways (see Section 5.2 for details on how they are considered by the optimizer):

- $R \rightarrow R$: A read followed by a read. The I/O for the second read may be eliminated using the in-memory copy of the shared block used to serve the first read (assuming there is no write of the same block in between). Alternatively, a new schedule may be able to reorder the reads, such that the first read in the original schedule becomes the second in the new schedule and saves its I/O. In either case, to realize the sharing opportunity, the shared block has to be kept in memory until the reuse.

- $W \rightarrow R$: A write followed by a read. The I/O for the read may be eliminated using the in-memory copy of the shared block used to serve the write (assuming here is no other write of the same block in between). To realize this opportunity, the shared block has to be kept in memory until the read. Unlike the case of $R \rightarrow R$, the new schedule cannot reorder these accesses because it would violate the $W \rightarrow R$ dependence.

- $W \rightarrow W$: A write followed by a write. The first write may be eliminated since it will be overwritten by the second one (assuming there is no other write of the same block in between). The shared data need not be kept in memory. Like the case of $W \rightarrow R$ but unlike $R \rightarrow R$, the new schedule cannot reorder these accesses because it would violate the $W \rightarrow W$ dependence.

Note that a $R \rightarrow W$ co-access does not make a sharing opportunity because neither the read nor the write can be saved.[4]

The resemblance between Definitions 2 and 3 is not a coincidence: two accesses to the same data may impose an execution order on the two statement instances and thus induce a dependence (if either access is a write), or may represent an opportunity for reducing I/O (if the co-access is not $R \rightarrow W$). However, their differences should also be clear. First, dependences capture the ordering constraints that must be preserved for any transformation, whereas sharing opportunities capture data reuse relationships that may potentially lead to I/O savings. Second, because of their distinct purposes, they stem from different subsets of co-access types: $R \rightarrow W$ can be a dependence but not a sharing opportunity, whereas $R \rightarrow R$ can be a sharing opportunity but not a dependence.

It is important to point out that the extent polyhedron of a co-access characterizes fine-grained, *instance-level* relationships, not a coarse-grained, *statement-level* relationship. Nonetheless, the polyhedron can be succinctly represented in an algebraic form (system of inequalities), instead of literal enumerations of integer points in the polyhedron. For example, in Example 1, $s_1 W C \rightarrow s_2 R C$ is both a dependence and a sharing opportunity, and $\mathbb{P}(s_1 W C \rightarrow s_2 R C) = \{(i, k, i', j', k') \mid i = i', k = k', 0 \leq i, i' < n_1, 0 \leq k, k' < n_2, 0 \leq j < n_3\}$. On the other hand, $s_2 R C \rightarrow s_1 W C$ is neither, because $\mathbb{P}(s_2 R C \rightarrow s_1 W C) = \emptyset$ (no instance of $s_2$ executes before any instance of $s_1$ in the original program).

It is also worth noting that *the arrow in $stA \rightarrow s't'A$ does not necessarily imply $s$ should textually precede $s'$ in the original program*. As a more dramatic example, consider the code below:

```
for (i=0; i<n; ++i) {
  A[i] = B[i];     // s1
  C[i] = A[n-1-i]; // s2
}
```

Two dependences (and sharing opportunities) with opposite directions exist at the same time, with polyhedra $\mathbb{P}(s_1 W A \rightarrow s_2 R A) =$ $\{(i, i') \mid i + i' = n - 1, 0 \le i \le (n-1)/2\}$ and $\mathbb{P}(s_2 \mathsf{R} A \to s_1 \mathsf{W} A) = \{(i', i) \mid i' + i = n - 1, 0 \le i' \le (n-2)/2\}$.

---

[4]Here we assume that a write operation does *not* include first reading that data. If a statement performs a read-modify-write, the read and the write are modeled as two separate accesses. This assumption still holds in the presence of disk blocks, because the unit of I/O is a logical array block as opposed to an individual array element.

## 5. THE OPTIMIZER

This section presents the design and implementation of our I/O sharing optimizer. At a high level, the optimizer translates individual data dependences and sharing opportunities, after necessary preprocessing, to constraints on schedules, which represent possible program transformations. The optimizer then uses an Apriori-like algorithm to efficiently enumerate feasible combinations of sharing opportunities while satisfying all dependences. Each feasible combination leads to a legal plan, which is then evaluated in terms of its memory requirement and total I/O cost. Finally, given the amount of memory available, the plan with the least I/O cost is chosen and converted into code for compilation and execution.

### 5.1 Preprocessing and Pruning

Before passing the sets of dependences and sharing opportunities on to the rest of the optimizer, we preprocess them by pruning out possibilities that either can be safely ignored by optimization, or need to be ignored to make optimization tractable. In this section, we describe two pruning techniques in these respective categories, and then discuss how extraction and preprocessing of dependences and sharing opportunities are implemented. Both pruning techniques stem from our concept of *linear sharing model*, which we first introduce below.

With any program schedule, every statement instance is executed at a specific time, which defines a linear ordering of all statement instances. I/O sharing effectively only happens between consecutive accesses to the same data in time order. To understand this statement, imagine that when a statement instance touches a shared piece of data, it becomes the owner of the data. A subsequent reuse of the data is always "charged" to the owner, and the new user becomes the new owner. We call this the *linear sharing model*. By considering sharing only between consecutive accesses, we avoid the problem of over-counting reuses. For example, consider three consecutive reads to the same data. There are only two pairs of consecutive accesses in time order, corresponding to two reuses. Including the non-consecutive accesses (the first and the third) would give us three, which is too much.

**No Write in Between** In light of the linear sharing model, the "no-write-in-between" rule states that, given a sharing opportunity $\mathfrak{a} \to \mathfrak{a}'$, any pair of statement instances $(\vec{x}, \vec{x}') \in \mathbb{P}(\mathfrak{a} \to \mathfrak{a}')$ can be removed from the polyhedron if there is a write to the same array block that executes between $\vec{x}$ and $\vec{x}'$ in the original program. This rule makes sense because, to preserve program semantics, no legal schedule can move the write before $\vec{x}$ or after $\vec{x}'$; hence, in no legal schedule will $\vec{x}$ and $\vec{x}'$ ever be consecutive accesses.

The no-write-in-between rule also applies to dependences: given a dependence $\mathfrak{a} \to \mathfrak{a}'$, any pair of statement instances $(\vec{x}, \vec{x}') \in \mathbb{P}(\mathfrak{a} \to \mathfrak{a}')$ can be removed from the polyhedron if there is a write to the same array block by some statement instance $\vec{y}$ that executes between $\vec{x}$ and $\vec{x}'$ in the original program. The rule is applicable in this setting because the ordering constraint between $\vec{x}$ and $\vec{x}'$ would be redundant, as it is implied by the constraints between $\vec{x}$ and $\vec{y}$, and between $\vec{y}$ and $\vec{x}'$.

**Multiplicity Reduction** We define the *multiplicity* of a sharing opportunity as follows. A sharing opportunity is *many-one* if each source instance is related to at most one target instance in the extent polyhedron, *one-many* if each target instance is related to at most one source instance, *one-one* if both, or *many-many* if neither. For a sharing opportunity that is not one-one, there exists an instance $\vec{x}$ related to multiple other instances. However, by the linear sharing model, only one of these instances can possibly form a real sharing relationship with $\vec{x}$. Ideally, the optimizer should explore all possibilities when realizing the sharing opportunity; however, doing so is impractical because it would blow up the search space. As a practical alternative, we perform a multiplicity reduction step to make all sharing opportunities one-one. Care is taken to minimize the impact on optimality. For details, see Remark A.1 in the appendix. Experiment results in Section 6 confirm that such reduction does not miss interesting solutions.

Note that multiplicity reduction is not applied to dependences, because a legal schedule must preserve the execution order of *all* statement instances in a dependence's polyhedron.

**Extracting and Preprocessing Dependences and Sharing Opportunities** We use isl [23], a library for manipulating integer points in polyhedra, for extracting program dependences. The algorithm used by isl was first introduced in [11]. Because of the similarity of dependences and sharing opportunities, we adapt the algorithm to extract sharing opportunities. The library supports removing transitively-covered dependent statement instances, which we use to produce no-write-in-between dependences and sharing opportunities. We then apply our multiplicity reduction algorithm on the sharing opportunities only. Henceforth, all dependences and sharing opportunities are assumed to be no-write-in-between; in addition, all sharing opportunities are assumed to be one-one. With a slight abuse of notation, we still use $\mathbb{P}(\mathfrak{a} \to \mathfrak{a}')$ to denote the polyhedron of a dependence or sharing opportunity after the aforementioned preprocessing.

### 5.2 Deriving Constraints

There are three types of constraints imposed on a schedule.

**Dimensionality Constraints** At the very least—not even considering data dependences and sharing opportunities—a legal schedule must map every statement instance in the original program to a unique execution time. This requirement can be satisfied by ensuring 1) instances belonging to the same statement are mapped to different times, and 2) any two instances belonging to different statements are mapped to different times. Below we explain how 1) can be translated into concrete constraints on the schedule. 2) is handled by the optimizer's search algorithm as it involves schedules of multiple statements and thus needs global coordination.[5]

A schedule $\Theta_s$ of statement $s$ is essentially a linear map, $\Theta_s : \mathbb{D}_s \mapsto \mathbb{Z}^{\tilde{d}+1}$ (see Section 4.2 for the definition of $\tilde{d}$ and $d_s$ used below). Ensuring all instances in $\mathbb{D}_s$ map to different images means $\Theta_s$ has to be injective. Thus, the null space of $\Theta_s$ should have dimension 0, i.e., $\dim \operatorname{null} \Theta_s = 0$. By the rank-nullity theorem in linear algebra, $\dim \mathbb{D}_s = \dim \operatorname{null} \Theta_s + \operatorname{rank} \Theta_s$, and the fact that $\dim \mathbb{D}_s = d_s$, we have $\operatorname{rank} \Theta_s = d_s$. In other words, the matrix representation of $\Theta_s$ should have exactly $d_s$ linearly independent rows out of the first $\tilde{d}$ rows (as the last dimension is a constant which does not contribute to the dimensionality).

The optimizer finds the matrix representation of $\Theta_s$ in a row-by-row fashion. When choosing each row, we use Algorithm 1 below to enumerate whether the current row should be linearly independent of previously found rows.

**Dependence Constraints** By definition, a schedule $\Theta = \{\Theta_s \mid s \in \mathcal{S}\}$ is legal only if for any dependence $\mathfrak{a} \to \mathfrak{a}', \forall (\vec{x}, \vec{x}') \in \mathbb{P}(\mathfrak{a} \to \mathfrak{a}'), \Theta_{\mathfrak{a},s}\vec{x} < \Theta_{\mathfrak{a}',s}\vec{x}'$. This < condition should not be confused with the one in the definition of extent polyhedron (Definition 1):

---

[5]The optimizer either assigns different constants for the last schedule dimensions of different statements, or tries to separate instances of different statements at an earlier dimension. Due to the space limit, we omit the technical details.

```
Algorithm 1: EnumRow(i,j,k)
  Input: Statement ID i, current row index j (1-based), and k, number of
         independent rows before row j
  Output: a list of Booleans indicating whether row j can be linear
          independent of previous rows
1 if d̃ − j = d_{s_i} − k then
2 |   return {1}
3 else
4 |   return {0, 1}
```

that condition is in terms of the *original* schedule of the program, whereas this condition applies to a *new* schedule. To translate the dependence constraint into a linear form, we first let

$$X_{s,s'}^q \triangleq (\theta_s^1 \vec{x} = \theta_{s'}^1 \vec{x}') \wedge \cdots \wedge (\theta_s^{q-1} \vec{x} = \theta_{s'}^{q-1} \vec{x}') \wedge (\theta_s^q \vec{x} < \theta_{s'}^q \vec{x}').$$

Then, by definition of $\prec$, we can write the dependence constraint as

$$\forall (\vec{x}, \vec{x}') \in \mathbb{P}(\mathfrak{a} \to \mathfrak{a}'), \; X_{\mathfrak{a}.s, \mathfrak{a}'.s}^1 \vee \cdots \vee X_{\mathfrak{a}.s, \mathfrak{a}'.s}^{\tilde{d}+1},$$

since $\prec$ can be satisfied at any depth. Note that the $X$ terms are mutually exclusive; i.e., only one of them can be true. If the $q$-th term is true, we say the dependence is *strongly satisfied* at depth $q$.

The next question is how to handle the quadratic-form constraints, i.e., vector inner products such as $\theta_s^q \vec{x}$. As a concrete example, let us examine the dependence $s_2 WE \to s_2 WE$ in Example 1. Its polyhedron is $\mathbb{P} = \{(i, j, k, i', j', k') \mid i' - i = 0, j' - j = 0, k' - k - 1 = 0\}$.[6] Suppose we want to find the constraint on a schedule dimension $q$ such that $\theta_{s_2}^q \cdot (i, j, k) < \theta_{s_2}^q \cdot (i', j', k')$. If we let $\theta_{s_2}^q = (\alpha, \beta, \gamma)$, the target constraint can be rewritten as $\alpha i + \beta j + \gamma k < \alpha i' + \beta j' + \gamma k'$, or $-\alpha i - \beta j - \gamma k + \alpha i' + \beta j' + \gamma k' - 1 \geq 0$. Note that this constraint is quadratic and thus does not directly fit in the polyhedral model. Fortunately, the following lemma provides a powerful mechanism to linearize such constraints.

**Lemma 1** (Affine Form of the Farkas Lemma [20]). *Let $\mathbb{P}$ be a nonempty polyhedron defined by $p$ affine inequalities:*

$$\vec{a}_k \vec{x} + \vec{b}_k \geq 0, \; k = 1, \ldots, p. \tag{1}$$

*Then $\forall \vec{x} \in \mathbb{P}, \vec{\theta} \vec{x} \geq 0$ iff there exist $\lambda_0, \ldots, \lambda_p \geq 0$ such that*

$$\vec{\theta} \vec{x} \equiv \lambda_0 + \sum_k \lambda_k (\vec{a}_k \vec{x} + \vec{b}_k).$$

Note that constraints of forms other than $\geq 0$ as in (1) can be rewritten so that the affine form of the Farkas Lemma applies. For example, $\theta_s^q \vec{x} < \theta_{s'}^q \vec{x}'$ can be rewritten as $\theta_{s'}^q \vec{x}' - \theta_s^q \vec{x} - 1 \geq 0$; an equality can be split into two inequalities, $\geq 0$ and $\leq 0$.

Continuing the above example, by Lemma 1, $-\alpha i - \beta j - \gamma k + \alpha i' + \beta j' + \gamma k' - 1 \equiv \lambda_0 + \lambda_1 (i' - i) + \lambda_2 (i - i') + \lambda_3 (j' - j) + \lambda_4 (j - j') + \lambda_5 (k' - k - 1)$. Comparing the coefficients of the iteration variables on both sides gives

$$-1 = \lambda_0 - \lambda_5, \; \alpha = \lambda_1 - \lambda_2, \; \beta = \lambda_3 - \lambda_4, \; \gamma = \lambda_5, \; \lambda_0, \ldots, \lambda_5 \geq 0.$$

By eliminating $\lambda_0, \ldots, \lambda_5$, we obtain $\alpha \in \mathbb{Z}, \beta \in \mathbb{Z}$, and $\gamma \geq 1$, which indeed preserve the execution order of dependent iterations.

A legal schedule should strongly satisfy each dependence in the program at a certain depth. As we have seen through the above example, the affine form of the Farkas Lemma helps us translate the condition of "strongly satisfying a dependence at a depth $q$" into a polyhedral constraint on the schedule's coefficients at dimension $q$. The union of such polyhedra for different depths characterizes the space of valid schedules for the statements involved in this particular dependence. The intersection of the results of these unions across all dependences characterizes the space of legal schedules for the entire program. Conceptually, the optimizer uses the dependence constraints to narrow the search space down to legal schedules only; practically, it employs a less expensive, depth-by-depth algorithm as discussed in Section 5.3.

---

[6] For the ease of presentation, we have omitted the parameter dimensions $n_1$, $n_2$, $n_3$ and the constant dimension; they are unimportant for the purpose of this example.

**Table 1:** Constraints on statement schedules $\Theta_s$ and $\Theta_{s'}$ that realize a sharing opportunity $\mathfrak{a} \to \mathfrak{a}'$. Here, $\mathbb{P} = \mathbb{P}(\mathfrak{a} \to \mathfrak{a}')$, $s = \mathfrak{a}.s$, and $s' = \mathfrak{a}'.s$.

| non-self ($s \neq s'$) | |
|---|---|
| W → R, W → W | $\exists c > 0, \forall (\vec{x}, \vec{x}') \in \mathbb{P}, \Theta_{s'} \vec{x}' - \Theta_s \vec{x} = (0, \ldots, 0, 0, c)$ |
| R → R | $\exists c \neq 0, \forall (\vec{x}, \vec{x}') \in \mathbb{P}, \Theta_{s'} \vec{x}' - \Theta_s \vec{x} = (0, \ldots, 0, 0, c)$ |
| self ($s = s'$) | |
| W → R, W → W | $\forall (\vec{x}, \vec{x}') \in \mathbb{P}, \Theta_s \vec{x}' - \Theta_s \vec{x} = (0, \ldots, 0, 1, 0)$ |
| R → R | $\exists c \in \{\pm 1\}, \forall (\vec{x}, \vec{x}') \in \mathbb{P}, \Theta_s \vec{x}' - \Theta_s \vec{x} = (0, \ldots, 0, c, 0)$ |

**Sharing Opportunity Constraints** Realizing a sharing opportunity in a schedule also imposes certain constraints on the schedule. To reduce the duration for which the shared data has to remain in memory, we require that related statement instances in a non-self sharing opportunity $\mathfrak{a} \to \mathfrak{a}'$ (where $\mathfrak{a}.s \neq \mathfrak{a}'.s$) be scheduled to times that differ only in the last constant time dimension. However, this requirement would not work for a self sharing opportunity $\mathfrak{a} \to \mathfrak{a}'$ (where $\mathfrak{a}.s = \mathfrak{a}'.s$), because related statement instances have the same constant for their last schedule dimension and thus enforcing it would schedule two instances to the same time. Thus, we instead require them to be scheduled to consecutive times, ignoring the last constant time dimension. In mathematical terms, realizing a sharing opportunity means satisfying the constraints listed in Table 1 according to its type. We special-case R → R sharing opportunities because a pair of related statement instances may have their execution order reversed by a new schedule without altering the original program semantics. Note that the polyhedra in Table 1 are after the preprocessing steps discussed in Section 5.1.

Each dimension (except the last one) of these constraints can easily be converted into linear constraints on the schedules by applying the affine form of the Farkas Lemma. The last constant dimension for all statements can be determined by a simple algorithm based on topological sort.

## 5.3 Search Algorithm

The goal of optimization is to find a schedule that minimizes I/O cost, or maximizes I/O savings, for a program given a certain amount of available memory. I/O savings come from the realization of sharing opportunities. It is important to note the following:

- Not all sharing opportunities can be realized simultaneously; some may be in direct conflict, as shown in Example 1.
- Maximizing the number of realized sharing opportunities does not necessarily minimize the total I/O cost, because the memory requirement may exceed the given cap, or because the amount of I/O saved by individual sharing opportunities varies depending on the sizes of array blocks and their iteration domains.

A naïve approach is to enumerate the power set of $O$, the set of all sharing opportunities, and for each candidate subset check if its member sharing opportunities can all be realized by a schedule while satisfying all dimensionality and dependence constraints. We propose a better algorithm based on the following key observation:

**Lemma 2** (Apriori Property). *If a set of sharing opportunities cannot be realized simultaneously, nor can any of its supersets.*

This lemma immediately suggests an algorithm similar to Apriori [1]. Algorithm 2 shows the details. The algorithm proceeds in the order of increasing size of sharing opportunity combinations. A set of $k$ sharing opportunities is considered a candidate only if all its subsets of size $k − 1$ are found to be feasible already (Line 5). A candidate is feasible only if it survives the *FindSchedule* test (details below), which attempts to find a schedule that realizes all the sharing opportunities in the candidate while satisfying all dimensionality and dependence constraints.

**Algorithm 2:** Apriori-like search.

**Input**: Set containing all sharing opportunities $O$, set containing all dependences $\mathcal{D}$
**Output**: Set of legal schedules, each satisfying a different combination of sharing opportunities
1  $C_1 \leftarrow \{o \mid o \in O, \text{FindSchedule}(\{o\}, \mathcal{D}) \neq \varnothing\}$
2  $T \leftarrow \{\text{FindSchedule}(\{o\}, \mathcal{D}) \mid o \in C_1\}$
3  $k \leftarrow 2$
4  **while** $C_{k-1} \neq \varnothing$ and $k \leq |O|$ **do**
5   $\quad C_k \leftarrow \{c \mid c \subseteq O, |c| = k, c$'s subsets of size $k-1$ all in $C_{k-1}\}$
6   $\quad$ **foreach** $c \in C_k$ **do**
7   $\quad\quad t \leftarrow \text{FindSchedule}(c, \mathcal{D})$
8   $\quad\quad$ **if** $t = \varnothing$ **then** $C_k \leftarrow C_k - \{c\}$ **else** $T \leftarrow T \cup \{t\}$
9   $\quad k \leftarrow k + 1$
10 **return** $T$

FindSchedule (Algorithm 3), repeatedly called by the search algorithm, searches for a legal schedule for a candidate sharing opportunity set. Because each dependence constraint can be satisfied at any of the $\tilde{d} + 1$ depths and it is computationally too expensive to consider all possibilities, the algorithm tries to satisfy each dependence constraint in a greedy fashion, from depth 1 to depth $\tilde{d} + 1$. Satisfying a dependence constraint at an early depth may lead to "over-separation" of statements (through enforcing the lexicographic order of statement instances) and prevent I/O sharing. To address this issue, we give higher priority to sharing opportunity constraints (Lines 13–26) and lower priority to dependence constraints (Lines 39–43). A similar greedy algorithm is used to satisfy dimensionality constraints (Lines 28–38). Note that Algorithm 3 involves many basic polyhedral operations, e.g., intersection and applying the affine form of the Farkas Lemma. The `isl` library [23] provides efficient implementation of these operations.

One subtlety is worth noting. For two feasible sharing opportunity sets $Q$ and $Q'$ where $Q \subset Q'$, FindSchedule may produce the same schedule. Indeed, a schedule satisfying the sharing opportunity constraints for $Q$ may happen to also satisfy those for $Q' \setminus Q$, because the algorithm does not explicitly consider other sharing opportunities when trying to satisfy $Q$. However, as discussed earlier, "accidentally" realizing more sharing opportunities may not be desirable, as it may increase memory requirement. Thus, when generating code for a schedule (Section 5.5), we consider the set $Q$ of sharing opportunities it is supposed to realize, and inject appropriate code to exploit only $Q$.[7] In any case, the search will eventually consider superset $Q' \supset Q$, so no good schedule will be missed.

## 5.4 Cost Evaluation

The search algorithm returns a list of legal schedules, each satisfying a particular subset of sharing opportunities. Next, we evaluate each schedule in terms of memory requirement and I/O cost.

**Memory Requirement** We want to compute the maximum amount of memory required by a schedule $\Theta$ found for a sharing opportunity set $Q$. First consider the baseline, where no sharing opportunities are realized. For a statement to successfully execute at time $\vec{\tau}$, all array

**Algorithm 3:** FindSchedule($Q, \mathcal{D}$).

**Input**: Sharing opportunity set $Q$, dependence set $\mathcal{D}$
1  $n \leftarrow |\mathcal{S}|$, the number of statements
2  $\tilde{d} \leftarrow \max_{s \in \mathcal{S}} d_s$
3  $Q_{sw} \leftarrow$ self sharing opportunities of types W → R, W → W
4  $Q_{sr} \leftarrow$ self sharing opportunities of type R → R
5  $Q_{nw} \leftarrow$ non-self sharing opportunities of types W → R, W → W
6  $Q_{nr} \leftarrow$ non-self sharing opportunities of type R → R
7  $k_1, \ldots, k_n \leftarrow 0; \Theta_1, \ldots, \Theta_n \leftarrow \varnothing$
8  Let $\theta^d$ denote $(\theta_1^d, \ldots, \theta_n^d)$, the $d$-th dimension of schedules
9  **for** $d \leftarrow 1$ **to** $\tilde{d}$ **do**
   // Initialize the space of schedules for dimension $d$
10  $\quad \mathbb{X}_d \leftarrow$ the polyhedron containing all integer points
    // Weakly satisfy *remaining* dependence constraints
11  $\quad$ **foreach** $\mathfrak{a} \to \mathfrak{a}' \in \mathcal{D}$ **do**
12  $\quad\quad \mathbb{X}_d \leftarrow \mathbb{X}_d \cap \{\theta^d \mid \forall (\vec{x}, \vec{x}') \in \mathbb{P}(\mathfrak{a} \to \mathfrak{a}'), \theta_{\mathfrak{a}'.s}^d \vec{x}' - \theta_{\mathfrak{a}.s}^d \vec{x} \geq 0\}$
    // Sharing opportunity constraints
13  $\quad$ **foreach** $\mathfrak{a} \to \mathfrak{a}' \in Q_{nw} \cup Q_{nr}$ **do**
14  $\quad\quad \mathbb{X}_d \leftarrow \mathbb{X}_d \cap \{\theta^d \mid \forall (\vec{x}, \vec{x}') \in \mathbb{P}(\mathfrak{a} \to \mathfrak{a}'), \theta_{\mathfrak{a}'.s}^d \vec{x}' - \theta_{\mathfrak{a}.s}^d \vec{x} = 0\}$
15  $\quad$ **if** $d < \tilde{d}$ **then**
16  $\quad\quad$ **foreach** $\mathfrak{a} \to \mathfrak{a}' \in Q_{sw} \cup Q_{sr}$ **do**
17  $\quad\quad\quad \mathbb{X}_d \leftarrow \mathbb{X}_d \cap$
18  $\quad\quad\quad \{\theta^d \mid \forall (\vec{x}, \vec{x}') \in \mathbb{P}(\mathfrak{a} \to \mathfrak{a}'), \theta_{\mathfrak{a}'.s}^d \vec{x}' - \theta_{\mathfrak{a}.s}^d \vec{x} = 0\}$
19  $\quad$ **else**
20  $\quad\quad$ **foreach** $\mathfrak{a} \to \mathfrak{a}' \in Q_{sw}$ **do**
21  $\quad\quad\quad \mathbb{X}_d \leftarrow \mathbb{X}_d \cap$
22  $\quad\quad\quad \{\theta^d \mid \forall (\vec{x}, \vec{x}') \in \mathbb{P}(\mathfrak{a} \to \mathfrak{a}'), \theta_{\mathfrak{a}'.s}^d \vec{x}' - \theta_{\mathfrak{a}.s}^d \vec{x} = 1\}$
23  $\quad\quad$ **foreach** $\mathfrak{a} \to \mathfrak{a}' \in Q_{sr}$ **do**
24  $\quad\quad\quad \mathbb{X}_d \leftarrow \mathbb{X}_d \cap$
25  $\quad\quad\quad (\{\theta^d \mid \forall (\vec{x}, \vec{x}') \in \mathbb{P}(\mathfrak{a} \to \mathfrak{a}'), \theta_{\mathfrak{a}'.s}^d \vec{x}' - \theta_{\mathfrak{a}.s}^d \vec{x} = -1\} \cup$
26  $\quad\quad\quad \{\theta^d \mid \forall (\vec{x}, \vec{x}') \in \mathbb{P}(\mathfrak{a} \to \mathfrak{a}'), \theta_{\mathfrak{a}'.s}^d \vec{x}' - \theta_{\mathfrak{a}.s}^d \vec{x} = 1\})$
27  $\quad$ **if** $\mathbb{X}_d = \varnothing$ **then return** $\varnothing$
    // Dimensionality constraints
28  $\quad$ **for** $i \leftarrow 1$ **to** $n$ **do**
29  $\quad\quad f \leftarrow false$
30  $\quad\quad$ **foreach** $l \leftarrow EnumRow(i, d-1, k_i)$ **do**
31  $\quad\quad\quad$ **if** $l = 0$ **then** $T \leftarrow$ space spanned by $\Theta_i$
32  $\quad\quad\quad$ **else** $T \leftarrow$ null space of $\Theta_i$
33  $\quad\quad\quad$ **if** $\mathbb{X}_d \cap T \neq \varnothing$ **then**
34  $\quad\quad\quad\quad \mathbb{X}_d \leftarrow \mathbb{X}_d \cap T$
35  $\quad\quad\quad\quad k_i \leftarrow k_i + l$
36  $\quad\quad\quad\quad f \leftarrow true$
37  $\quad\quad\quad\quad$ **break**
38  $\quad\quad$ **if** $f = false$ **then return** $\varnothing$
    // Strongly satisfy remaining dependence constraints
39  $\quad$ **foreach** $\mathfrak{a} \to \mathfrak{a}' \in \mathcal{D}$ **do**
40  $\quad\quad T \leftarrow \{\theta^d \mid \forall (\vec{x}, \vec{x}') \in \mathbb{P}(\mathfrak{a} \to \mathfrak{a}'), \theta_{\mathfrak{a}'.s}^d \vec{x}' - \theta_{\mathfrak{a}.s}^d \vec{x} > 0\}$
41  $\quad\quad$ **if** $\mathbb{X}_d \cap T \neq \varnothing$ **then**
42  $\quad\quad\quad \mathbb{X}_d \leftarrow \mathbb{X}_d \cap T$
43  $\quad\quad\quad \mathcal{D} \leftarrow \mathcal{D} - \{\mathfrak{a} \to \mathfrak{a}'\}$
44  $\quad \theta_1^d, \ldots, \theta_n^d \leftarrow$ sample a point from $\mathbb{X}_d$
45  $\quad$ **foreach** $i \leftarrow 1$ **to** $n$ **do** $\Theta_i \leftarrow \Theta_i \cup \{\theta_i^d\}$
46 Find constants for the last dimensions of $\Theta_1, \ldots, \Theta_n$
47 **return** $\Theta = \{\Theta_1, \ldots, \Theta_n\}$

blocks it accesses must all be in memory at $\vec{\tau}$. Therefore, the baseline memory requirement at time $\vec{\tau}$, $M(\vec{\tau})$, can be computed by first finding the iteration instance $\vec{x} = \Theta^{-1}(\vec{\tau})$, and then summing up all the sizes of the array blocks $\vec{x}$ accesses. Each realized sharing opportunity except those of type W → W can require additional memory for keeping the shared array block until the reuse occurs. For each sharing opportunity $\mathfrak{a} \to \mathfrak{a}' \in Q$, for each $(\vec{x}, \vec{x}') \in \mathbb{P}(\mathfrak{a} \to \mathfrak{a}')$, the shared array block $\mathfrak{a}.A[\mathfrak{a}.\Phi \vec{x}]$ has to be kept in memory between time $\Theta_{\mathfrak{a}.s} \vec{x}$ and $\Theta_{\mathfrak{a}'.s} \vec{x}'$. Thus we can find for each time $\vec{\tau}$ all the

---
[7]This brings up the point that a schedule alone does not completely dictate what and how I/O sharing is achieved; it only specifies the execution timing of statement instances, a necessary but not sufficient condition for sharing. Code generation must ensure that appropriate I/O and memory buffer management actions are taken to enable sharing.

771

additional array blocks that have to be in memory at that time, and add their sizes to $M(\vec{t})$. Finally, taking the maximum of $M(\vec{t})$ across all $\vec{t}$'s gives the memory requirement of the schedule.

**I/O Cost** We adopt a simple I/O cost model that predicts the total I/O time as a linear function of the total read and write volumes (in the number of bytes). More refined models (e.g., charging an overhead for each I/O request) can be easily incorporated, though as shown by the experiments in Section 6, our simple model already provides very accurate estimates (thanks to our framework's ability to capture instance-level I/O sharing).

Without realizing any sharing opportunity, the baseline I/O cost for a statement can be computed by summing up the sizes of all array blocks it accesses over its iteration domain. Realized I/O sharing opportunities involving the statement can save some of the baseline I/O operations and cut down the cost. For example, for a sharing opportunity $\mathfrak{a} \to \mathfrak{a}'$ where $\mathfrak{a}' = \langle s', \mathsf{R}, A, \Phi' \rangle$, at any target iteration $\vec{x}' \in \{\vec{x}' \mid (\vec{x}, \vec{x}') \in \mathbb{P}(\mathfrak{a} \to \mathfrak{a}')\}$ a read of array block $A[\Phi' \vec{x}']$ is saved; for one where $\mathfrak{a} = \langle s, \mathsf{W}, A, \Phi \rangle$ and $\mathfrak{a}' = \langle s', \mathsf{W}, A, \Phi' \rangle$, at any source iteration $\vec{x} \in \{\vec{x} \mid (\vec{x}, \vec{x}') \in \mathbb{P}(\mathfrak{a} \to \mathfrak{a}')\}$ a write of array block $A[\Phi \vec{x}]$ is saved. With a union operation across all realized sharing opportunities, we can find the I/O savings for every iteration instance. Summing up all the savings over all iteration instances gives the total I/O savings of the given schedule; subtracting it from the baseline I/O cost gives the actual I/O cost of the schedule.

**Remark** All computation above happens in a symbolic, algebraic fashion. A schedule's memory requirement and I/O cost are represented as polynomials (piecewise quasipolynomials to be exact) in the global parameters $\vec{p}$. The advantage of this approach is that schedule search and evaluation need to be done only once for a given program "template"; should the parameters (array and block sizes) change, we can simply plug the new values into the polynomials instead of performing optimization all over again.

## 5.5 Code Generation

The schedule chosen by the optimizer is subsequently transformed into C code with `for` and `if` control structures for compilation and execution. This process is the reverse of the program analysis that happens before optimization: program analysis extracts a polyhedral representation from code, while code generation converts the optimized polyhedral representation back to code. Recent advances in polyhedral compiler construction have produced efficient code generation tools such as CLooG [3, 22], which is incorporated in production compilers such as GCC and also used by us.

As a concrete example, let us see what code is generated for the program in Example 1. Suppose the following three sharing opportunities are satisfied: $s_1 \mathsf{W} C \to s_2 \mathsf{R} C$, $s_2 \mathsf{W} E \to s_2 \mathsf{R} E$, $s_2 \mathsf{W} E \to s_2 \mathsf{W} E$. Note that whether this yields the optimal I/O cost depends on the values of the parameters; here we use the parameterized schedule to demonstrate code generation only. The optimizer produces the following schedule for this set of sharing opportunities: $\Theta_{s_1} \vec{x}_{s_1} = (0, -i, k, 0)$, $\Theta_{s_2} \vec{x}_{s_2} = (j, -i, k, 1)$. The generated code for this schedule is listed below. It is easy to verify that this is equivalent to the hand-generated solution in Figure 1(b).[8]

```
for (i=-n1+1; i<=0; i++)
  for (k=0; k<=n2-1; k++) {
    C[-i,k]  = A[-i,k] + B[-i,k]; // s1
    E[-i,0] += C[-i,k] * D[k, 0]; // s2
  }
for (j=1; j<=n3-1; j++)
  for (i=-n1+1; i<=0; i++)
    for (k=0; k<=n2-1; k++)
      E[-i,j] += C[-i,k] * D[k,j]; // s2
```

---

[8] $C$ is not written in the case ($n_3 = 1$) shown in Figure 1(b). Although not reflected in the code shown here, our optimizer and execution engine check the value of $n_3$ and decide if $C$ needs to be written to disk.

For brevity, the code above does not contain explicit I/O operations. In practice, RIOTShare injects additional code to ensure that all array block accesses are fulfilled either by blocks already buffered in memory or by I/O (and displacing appropriate buffered blocks when necessary); the details are omitted. In general, RIOTShare relieves the burden of manually managing I/O from library and application developers.

## 6. EXPERIMENTS

**Setup** All our experiments were run on a desktop computer with an Intel Core i7-2600 four-core CPU, 8GB of memory, and a WD Caviar Black 7200RPM hard drive, running Ubuntu Linux 11.10. I/O and CPU time of plan execution was collected using the systemtap instrumentation tool. We verified that instrumentation overhead was negligible. To make it easier to understand results, we used the `ext2` file system, as it does not have journaling that would necessarily complicate result interpretation. To make I/O measurements meaningful, we turned off file system caching using the `O_DIRECT` flag when opening files. Under this setting, we benchmarked the I/O rates of the hard drive and found that sustained reads and writes were 96MB/s and 60MB/s, respectively. These numbers were used by the optimizer to convert the predicted I/O volume of plans to estimated I/O time. The in-core computation of tested programs was done by calling GotoBLAS2 [15], an optimized implementation of BLAS which is able to utilize all four cores on our machine.

**Storage Scheme** In the following experiments, matrices are stored in large, logical blocks. The blocks are laid out on disk in column-major order, and so are the elements within each block. Since every element in a matrix has a predetermined storage position, its index (row and column numbers) is not stored. This is a highly efficient storage scheme for dense matrices. We use our previously developed storage library, RIOTStore [26], for managing the storage and performing I/O. RIOTStore implements the LAB-tree (Linearized Array B-tree) and the DAF (Directly Addressable File) storage formats, both of which provide the storage scheme we want and work virtually identically for dense matrices.

**A Note on Optimization Time** The optimization time for all experiments below are reasonably short: 0.6 second for the matrix addition and multiplication program in Section 6.1, 2.1 seconds for the two matrix multiplications in Section 6.2, and 156.7 seconds (more on this next) for the linear regression program in Section 6.3. Even though it optimizes at the instance level, our optimizer avoids enumerating all statement instances by working with polyhedra. Hence, the complexity of optimization depends on the complexity of the program (such as the number of statements and dimensionalities of iteration domains) instead of the size of the data it operates on or the number of iterations each loop takes. This property can be seen directly from the algorithms in Section 5, and has been further confirmed by experiments on datasets of different scales.

The optimization time for the linear regression experiment is longer, because there are 7 operators (statements) and 16 sharing opportunities, and also because our optimizer is implemented in Python and single-threaded—we expect a multithreaded C implmentation to be significantly more efficient. Nevertheless, our optimizer is able to prune away 94% of the search space. As we shall see later, the optimization overhead is dwarfed by the I/O savings. Moreover, this overhead is not affected by the size of the dataset; it becomes more negligible when the dataset is larger. If needed, we can further improve optimization time for larger, more complex programs by *localizing optimization* to the most expensive code fragments, and by *combining plan enumeration and costing* so we can terminate search early as soon as acceptable plans are found.

**Datasets of Different Scales** We have run the following experiments with datasets of different scales and found consistent results. Also, as expected, optimization time for the same program does not change with the scale of the dataset. Due to the space limit, below we present only results on the largest dataset tested.

## 6.1 Matrix Addition and Multiplication

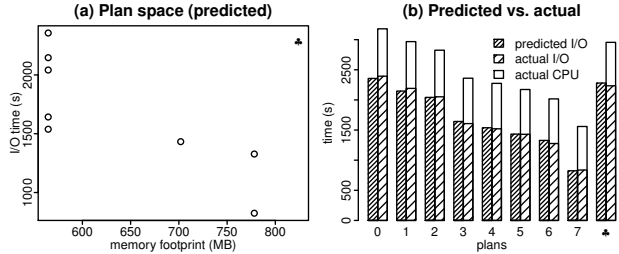

Figure 3: Matrix addition and multiplication: all plans.

We first test our optimizer with the program shown in Example 1. It consists of a matrix addition followed by a matrix multiplication. The actual matrix sizes used in this experiment are listed in Table 2. The optimizer finds 8 legal execution plans (including the unmodified original schedule). We plot these plans in Figure 3.

Table 2: Matrix addition and multiplication: matrix sizes.

| Matrix | Block size | # Blocks | Total size |
|---|---|---|---|
| $A, B, C$ | $6000 \times 4000$ | $12 \times 12$ | 25.6GB |
| $D$ | $4000 \times 5000$ | $12 \times 1$ | 1.8GB |
| $E$ | $6000 \times 5000$ | $12 \times 1$ | 2.7GB |

Figure 3(a) shows each plan's memory footprint and I/O time as estimated by the optimizer. The circles (∘) represent the 8 plans considered by the optimizer (the ♣ is explained below). We notice that a plan's memory footprint can only take one of three possible values, because there are limited combinations of which matrices' blocks to keep in memory. Among plans with identical memory footprint, all have different I/O costs. The plan with the lowest I/O cost, Plan 7 in the lower right corner, takes 836 seconds, while the original plan, Plan 0 in the upper left corner, takes 2394 seconds. The code generated from Plan 7 is equivalent to the one shown in Figure 1(b) for the general case. Since $n_3 = 1$ (the number of blocks in the second dimension of $D$ and $E$) in this experiment, Plan 7 is also effectively equivalent to the special case solution in Figure 1(a). Plan 7 satisfies three sharing opportunities: $s_1WC \rightarrow s_2RC$, $s_2WE \rightarrow s_2RE$ and $s_2WE \rightarrow s_2WE$. Note that because $n_3 = 1$, sharing opportunity $s_2RC \rightarrow s_2RC$ does not exist.

One may argue the comparison between Plan 0 and 7 is not fair: Plan 0 underutilizes the memory and could reduce I/O by buffering more data. Extra memory can be used to support sophisticated I/O sharing schedules as Plan 7, or to simply allow bigger array blocks. Which approach is better? To answer this question, we took Plan 0 and increased the number of rows in $A$, $B$, $C$, and $E$'s blocks from 6000 to 9000, and plotted it (♣) also in Figure 3(a). This modified plan consumes more memory than Plan 7, but still incurs far more I/O cost than it. This shows that blindly enlarging array blocks is not the best way of utilizing extra memory; cost-driven optimization like ours can give much better results.

Figure 3(b) compares the optimizer-predicted I/O cost with the actual I/O cost of executing the plan. This comparison shows our optimizer is impressively accurate in estimating the the I/O cost of plans; the average error is merely 1.7%. This high accuracy should be no surprise, because our optimizer is *precise* down to instance-level sharing and can calculate the exact number and amount of I/O. The only source of error is the simple I/O cost model we employ for predicting the I/O time from the amount of I/O; however, the error is small and does not affect optimization decisions.

Figure 3(b) also breaks the actual execution time of each plan down into CPU and I/O time. Because our optimizer only optimizes I/O, the CPU time is the same across all plans. With or without optimization, the program remains I/O-dominant. Therefore, maximizing I/O sharing brings the total execution time from the original 3180 seconds down to 1560 seconds, a 50.9% improvement.

**Comparing to Matlab and SciDB** Matlab and SciDB [8] represent state-of-the-art scientific computing and scientific data management systems, respectively. However, neither has RIOTShare's I/O sharing optimization capabilities. We have tested them using the same input program (properly translated) and the same data (properly converted and loaded). As with RIOTShare, both systems are allowed to use all four CPU cores of the machine. Running the test program without blocking in Matlab immediately gives a not-enough-memory error, due to the large data size. With blocking, Matlab's running time is 2.65 times that of our best plan. This suggests that not only is Matlab unable to optimize I/O, but it also has considerable control and storage overhead. Manually implementing our best plan in Matlab makes a big difference—the performance becomes 6% better than ours. The minor advantage may come from Matlab's higher in-memory math performance. This demonstrates that the ideas developed in this paper is readily transferable to existing systems and have great, platform-independent potential.

Although given a larger memory usage, SciDB takes 33.08 times more time than our best plan. This could be a result of not using a BLAS library or using an unoptimized one, and also not sharing I/O. While SciDB focuses on parallelization, its handling of execution on a single node seems to leave a big room for improvement on I/O efficiency.

## 6.2 Two Matrix Multiplications

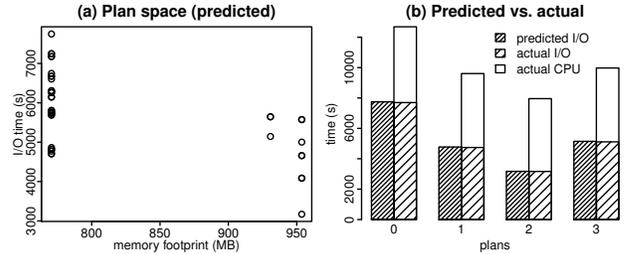

Figure 4: Two matrix multiplications: Config A, selected plans.

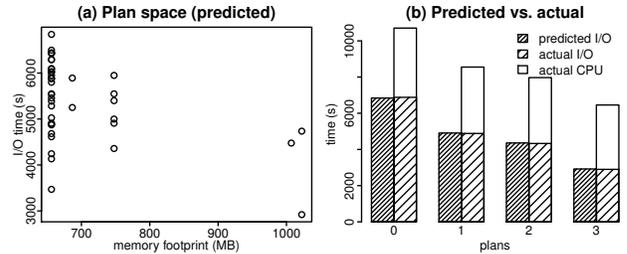

Figure 5: Two matrix multiplications: Config B, selected plans.

We next present results on a program with slightly more control structures and a larger search space. Matrix multiplication is a fundamental building block of many data analysis algrithms and routines and has long been the target of optimization efforts by researchers and HPC vendors. It is common to have multiple matrix multiplications in the same program. In the following, we consider two matrix multiplications, $C = AB$, $E = AD$, to be executed together. There are 9 sharing opportunities. We have tested this program with two different matrix size configurations; relevant information is summarized in Table 3 below.

```
for (i=0; i<n1; ++i)
  for (j=0; j<n2; ++j)
    for (k=0; k<n3; ++k)
      C[i,j] += A[i,k] * B[k,j]; // s1
for (i=0; i<n1; ++i)
  for (j=0; j<n4; ++j)
    for (k=0; k<n3; ++k)
      E[i,j] += A[i,k] * D[k,j]; // s2
```

**Table 3: Two matrix multiplications: matrix sizes.**

| Configuration | Matrix | Block size | # Blocks | Total size |
|---|---|---|---|---|
| A (Figure 4) | A | 8000 × 7000 | 6 × 6 | 15.2GB |
|  | B, D | 7000 × 3000 | 6 × 10 | 9.2GB |
|  | C, E | 8000 × 3000 | 6 × 10 | 10.8GB |
| B (Figure 5) | A | 2000 × 8000 | 18 × 6 | 12.8GB |
|  | B | 8000 × 6000 | 6 × 4 | 8.4GB |
|  | C | 2000 × 6000 | 18 × 4 | 6.4GB |
|  | D | 8000 × 7000 | 6 × 4 | 10.0GB |
|  | E | 2000 × 7000 | 18 × 4 | 7.6GB |

Under both configurations, the optimizer produced 40 plans. For the sake of presentation, we select four plans for demonstration below. Plan 0 enables no sharing opportunities; Plan 1 enables $s_1 \mathsf{W} C \to s_1 \mathsf{R} C$, $s_1 \mathsf{W} C \to s_1 \mathsf{W} C$, $s_2 \mathsf{W} E \to s_2 \mathsf{R} E$, and $s_2 \mathsf{W} E \to s_2 \mathsf{W} E$; Plan 2 enables all that Plan 1 enables, plus $s_1 \mathsf{R} A \to s_2 \mathsf{R} A$; Plan 3 enables $s_1 \mathsf{R} A \to s_2 \mathsf{R} A$, $s_1 \mathsf{R} B \to s_1 \mathsf{R} B$, and $s_2 \mathsf{R} D \to s_2 \mathsf{R} D$. Intuitively, Plan 1 uses two separate loop nests to accumulate $C$ and $E$ blocks in memory. Plan 2 in addition merges the two loop nests and shares the read of $A$. Plan 3 shares the I/O to $B$ and $D$ instead of $C$ and $E$. Each plan works the same way for both configurations, except with different size parameters.

Figure 4 and 5 summarize the plan spaces and characteristics of the selected plans. Figure 4(a) and Figure 5(a) clearly illustrate that different matrix size configurations have dramatic impact on plan cost and optimality. This observation highlights the need for automatic and systematic optimization, because code manually optimized based on expert knowledge or past experience has fragile performance. Even if one knows the best plan for the current problem, a slight change in memory cap or problem size can easily render the current solution inappropriate. The plans shown in Figure 4(b) and Figure 5(b) exemplify this observation. Plan 2 has the lowest I/O cost under Configuration A, but is suboptimal under Configuration B, where Plan 3 is the best. Comparing predicted and actual I/O times, we find the average error to be merely 0.6%. Even though matrix multiplication is traditionally considered CPU-dominant, the I/O and CPU time breakdown here actually reveals that, for big data, I/O is equally (if not more) expensive than CPU; optimizing I/O therefore provides good overall performance improvement.

### 6.3 Linear Regression: A Complete Program

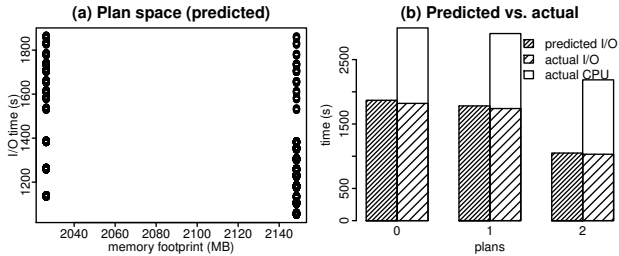

**Figure 6: Linear regression: selected plans.**

We next test RIOTShare with a commonly used statistical method— linear regression. Suppose we want to fit a linear model for a set of response variables $y = (y_1, \ldots, y_k)$ from $m$ predictor variables $x = (x_1, \ldots, x_m)$, i.e., $y_j = x'\beta_j + \epsilon_j$, where $\epsilon_j \sim \mathcal{N}(0, \sigma_j^2)$. Suppose $n$ i.i.d. observations are collected: $\{y_i, x_i\}_{i=1}^n$. Using the ordinary least square method, we can estimate the coefficient vectors $\beta_j$, which are column-combined to form a matrix $\beta$, *simultaneously* by $\hat{\beta} = (X'X)^{-1}X'Y$, where $X$ is formed by row-stacking $x_i$'s and $Y$ by row-stacking $y_i$'s. After obtaining $\hat{\beta}$, we further compute the Residual Sum of Squares: $RSS(Y_j - X\hat{\beta}_j) = \sum_{i=1}^n (y_{ji} - x'_i\hat{\beta}_j)^2$. Written in matrix form, this program has 7 steps (statements): $U = X'X$; $V = X'Y$; $W = U^{-1}$; $\hat{\beta} = WV$; $\hat{Y} = X\hat{\beta}$; $E = Y - \hat{Y}$; $R = RSS(E)$. Normally the number of response and predictor variables $k$ and $m$ are small but the number of observations $n$ can be very large. Below we consider a case where $k = 400$, $m = 4000$ and $n = 1.5 \times 10^6$. Table 4 summarizes the detailed size configuration of the matrices.

**Table 4: Linear regression: matrix sizes.**

| Matrix | Block size | # Blocks | Total size |
|---|---|---|---|
| $X$ | 60000 × 4000 | 25 × 1 | 44.7GB |
| $Y, \hat{Y}, E$ | 60000 × 400 | 25 × 1 | 4.5GB |
| $U, W$ | 4000 × 4000 | 1 × 1 | 122.1MB |
| $V, \hat{\beta}$ | 4000 × 400 | 1 × 1 | 12.2MB |

We have implemented the above linear regression program for optimization by RIOTShare. The input program has a sequence of 7 loop nests, one for each step. Following the design of BLAS, matrix transpose is not modeled as a separate operator, but as a flag passed to operations such as matrix multiplication.

Figure 6(a) plots all the plans generated by the optimizer. The best plan (bottom-right, Plan 2) uses only 6.0% more memory than the original unoptimized plan (top-left, Plan 0), but saves I/O time by 43.8%. This improvement comes from sharing the reads of $X$ for the two out-of-core matrix multiplications and eliminating the materialization of intermediate results. Figure 6 plots the predicted and actual running time of the two plans together with another Plan 1, which merely keeps $U$ and $V$ in memory during the multiplication. Again, the prediction is highly accurate, with maximum error 2.3%. In terms of total running time, the best plan gives 27.0% improvement over the original plan.

## 7. CONCLUSION AND FUTURE WORK

Big data analytics are often I/O-intensive. In this paper, we have presented RIOTShare, for representing and optimizing I/O patterns in such tasks. Building on the polyhedral model, RIOTShare strikes the balance between feasibility and flexibility of representation and optimization, by exploring the middle ground between the high-level, database-style operator-based query optimization and low-level, compiler-style loop-based code optimization. Experiments show that RIOTShare produces accurate estimates for the I/O costs of candidate plans, and finds nontrivial plans with significant I/O improvement under memory constraints.

In ongoing work, we are extending RIOTShare with the ability of selecting optimal array block sizes for storage. By jointly optimizing array block sizes and I/O sharing, the optimizer can produce better plans that use memory more effectively and result in more I/O savings. Since the polyhedral model is quite general, it would also be interesting to investigate the effectiveness of RIOTShare on programs involving a mix of array and matrix operations and database- or Pig-style operations.

# APPENDIX
## A. ADDITIONAL REMARKS

**Remark A.1 (Multiplicity Reduction; Section 5.1)** Multiplicity reduction is the process of tailoring non-one-one sharing opportunities to make them fit the linear sharing model. First note that a "many" side can only be a read access, because if a statement instance $\vec{x}$ is related to multiple instances with a write, only the instance closest to $\vec{x}$ in execution time forms a real sharing opportunity with $\vec{x}$ due to the *no write in between* rule, in which case the "many" side is not really a "many" side.

The algorithm we use works as follows. If a sharing opportunity is one-many or many-one, we reduce the multiplicity of the "many" side to "one". Specifically, we keep for any instance on the "one" side only the instance closest to it in execution time on the "many" side. Such reduction does not reduce the amount of I/O savings for this sharing opportunity, because each instance from the original "one" side can share I/O with at most one instance from the "many" side anyway, by the linear sharing model.

For many-many sharing opportunities, we first reduce them to many-one and then apply the reduction described above. The reduction from many-many to many-one cannot use the same idea as above, however; the potential problem is illustrated in Figure 7(a). If we keep for each source instance the target instance closest to it in execution time, many target instances may be ignored and thus reduce the amount of potential I/O savings. We solve this problem by ensuring the *rank*, or degree of freedom in the iteration variables, of both sides after the reduction do not decrease below the minimum of the original ranks of both sides. Continuing with the above example, suppose originally both sides are depth-1 loops: `for (i=0; i<2; ++i)`, both with rank 1 because `i` is a free variable. After the first reduction in Figure 7(a), the target side has constraint `i=0`, which means the rank decreases to 0 and renders the reduction invalid. Our algorithm carefully adds rank-preserving equality constraints and would produce a result as shown in Figure 7(b).

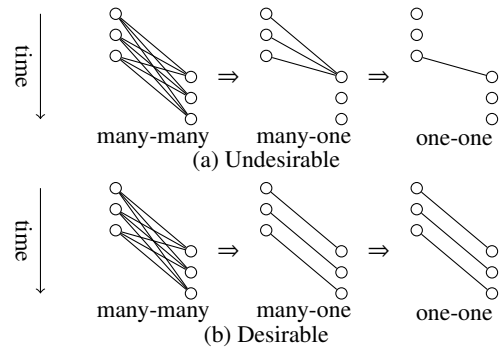

**Figure 7: Intricacy of multiplicity reduction for many-many.**

Note that the multiplicity reduction problem can be cast into a maximum bipartite matching problem and solved in $O(|V||E|)$ time. However, in our case, $|V|$ corresponds to the size of the iteration domain, which is *symbolic* and usually takes a big value, making regular algorithms inapplicable. Our algorithm, in contrast, works in $O(d_i d_j)$ time, where $d_i$ and $d_j$, source and target loop nest depths, are very small constants.

Although our multiplicity reduction algorithm does not reduce the amount of I/O sharing for each sharing opportunity, it is still possible that the reduced sharing opportunity cannot be satisfied with others at the same time while the original one can. This is expected in order to reduce the exponential search space and make optimization tractable.